\documentclass[letterpaper,11pt]{article}

\usepackage[latin1]{inputenc}
\usepackage[T1]{fontenc}
\usepackage{graphicx}
\usepackage{amsfonts}
\usepackage{amssymb}
\usepackage{amsmath} 
\usepackage{amsthm}
\usepackage{url}
\usepackage{rotating}

\usepackage{natbib}

\usepackage[margin=1in]{geometry}

\begin{document}

\title{Virtual screening of GPCRs: an \emph{in silico} chemogenomics approach}

\author{
Laurent Jacob\footnote{The first two authors contributed equally to
  this work}\\
Institut Curie, Paris, F-75248 France\\
INSERM, U900, Paris, F-75248 France\\
Ecole des Mines de Paris F-77300 France\\
\texttt{laurent.jacob@ensmp.fr}
\and
Brice Hoffmann\\
Institut Curie, Paris, F-75248 France\\
INSERM, U900, Paris, F-75248 France\\
Ecole des Mines de Paris F-77300 France\\
\texttt{brice.hoffmann@ensmp.fr}
\and
V\'eronique Stoven\\
Institut Curie, Paris, F-75248 France\\
INSERM, U900, Paris, F-75248 France\\
Ecole des Mines de Paris F-77300 France\\
\texttt{veronique.stoven@ensmp.fr}
\and
Jean-Philippe Vert\\
Institut Curie, Paris, F-75248 France\\
INSERM, U900, Paris, F-75248 France\\
Ecole des Mines de Paris F-77300 France\\
\texttt{jean-philippe.vert@ensmp.fr}
}

\maketitle

\begin{abstract}
  The G-protein coupled receptor (GPCR)
  superfamily is currently the largest class of therapeutic targets.
  \textit{In silico} prediction of interactions between GPCRs and
  small molecules is therefore a crucial step in the drug discovery
  process, which remains a daunting task due to the difficulty to
  characterize the 3D structure of most GPCRs, and to the limited
  amount of known ligands for some members of the superfamily.
  Chemogenomics, which attempts to characterize interactions between
  all members of a target class and all small molecules
  simultaneously, has recently been proposed as an interesting
  alternative to traditional docking or ligand-based virtual screening
  strategies.
  
  We propose new methods for \emph{in silico} chemogenomics and
  validate them on the virtual screening of GPCRs.  The methods
  represent an extension of a recently proposed machine learning
  strategy, based on support vector machines (SVM), which provides a
  flexible framework to incorporate various information sources on the
  biological space of targets and on the chemical space of small
  molecules. We investigate the use of 2D and 3D descriptors for small
  molecules, and test a variety of descriptors for GPCRs.  We show fo
  instance that incorporating information about the known hierarchical
  classification of the target family and about key residues in their
  inferred binding pockets significantly improves the prediction
  accuracy of our model. In particular we are able to predict ligands
  of orphan GPCRs with an estimated accuracy of $78.1\%$.
\end{abstract}

\section{Introduction}

The G-protein coupled receptor (GPCR) superfamily is comprised of an
estimated 600-1,000 members and is the largest known class of
molecular targets with proven therapeutic value. They are ubiquitous
in our body, being involved in regulation of every major mammalian
physiological system~\citep{Bockaert1999Molecular}, and play a role in
a wide range of disorders including allergies, cardiovascular
dysfunction, depression, obesity, cancer, pain, diabetes, and a
variety of central nervous system
disorders~\citep{Deshpande2006Targeting,Hill2006G-protein-coupled,Catapano2007G}.
They are integral membrane proteins sharing a common global topology
that consists of seven transmembrane alpha helices, an intracellular
C-terminal, an extracellular N-terminal, three intracellular loops and
three extracellular loops. There are four main classes of GPCRs (A, B,
C and D) depending on their sequence
similarity~\citep{Horn2003GPCRDB}. Their location on the cell surface
makes them readily accessible to drugs, and $30$ GPCRs are the targets
for the majority of best-selling drugs, representing about $40\%$ of
all prescription pharmaceuticals on the
market~\citep{Fredholm2007G-protein-coupled}. Besides, the human
genome contains several hundred unique GPCRs which have yet to be
assigned a clear cellular function, suggesting that they are likely to
remain an important target class for new drugs in the
future~\citep{Lin2004Orphan}.

Predicting interactions \textit{in silico} between small molecules and
GPCRs is not only of particular interest for the drug industry, but
also a useful step for the elucidation of many biological process.
First, it may help to decipher the function of so-called
\textit{orphan} GPCRs, for which no natural ligand is known. Second,
once a particular GPCR is selected as a target, it may help in the
selection of promising molecule candidates to be screened \textit{in
  vitro} against the target for lead identification.

\textit{In silico} virtual screening of GPCRs with classical
approaches is however a daunting task for at least two reasons. First,
the 3D structures are currently known for only two GPCRs (bovine
rhodopsin and human $\beta_2$-adrenergic receptor). Indeed, GPCRs,
like other membrane proteins, are notoriously difficult to
crystallize. As a result, docking strategies for screening small
molecules against GPCRs are often limited by the difficulty to model
correctly the 3D structure of the target. To circumvent the lack of
experimental structures, various studies have used 3D structural
models of GPCRs built by homology modeling using bovine rhodopsin as a
template structure. Docking a library of molecules into these modeled
structures allowed the recovery of known
ligands~\citep{Evers2005Structure-based}, and even identification of
new ligands~\citep{Cavasotto2003Structure-based}. However, docking
methods still suffer from docking and scoring inaccuracies, and
homology models are not always reliable-enough to be employed in
target-based virtual screening.  Methods have been proposed to enhance
the quality of the models by global optimization and flexible
docking~\citep{Cavasotto2003Structure-based}, or by using different
sets of receptor models. Nevertheless, these methods are expected to
show limited performances for GPCRs sharing low sequence similarity
with rhodopsin, especially in the case of receptors belonging to
classes B, C and D. Alternatively, ligand-based strategies, also known
as quantitative structure-activity relationship (QSAR), attempt to
predict new ligands from previously known ligands, often using
statistical or machine learning approaches.  Ligand-based approaches
are interesting because they do not require the knowledge of the
target 3D structure and can benefit from the discovery of new ligands.
However, their accuracy is fundamentally limited by the amount of
known ligands, and degrades when few ligands are known. Although these
methods were successfully used to retrieve strong GPCR
binders~\citep{Rolland2005G-protein-coupled}, they are efficient for
lead optimization within a previously identified molecular scaffold,
but are not appropriate to identify new families of ligands for a
target. At the extreme, they cannot be pursued for the screening of
orphan GPCRs.

Instead of focusing on each individual target independently from other
proteins, a recent trend in the pharmaceutical industry, often
referred to as \textit{chemogenomics}, is to screen molecules against
several targets of the same family
simultaneously~\citep{Kubinyi2004Chemogenomics,2006Chemical}. This
systematic screening of interactions between the chemical space of
small molecules and the biological space of protein targets can be
thought of as an attempt to fill a large 2D \emph{interaction matrix},
where rows correspond to targets, columns to small molecules, and the
$(i,j)$-th entry of the matrix indicates whether the $j$-th molecule
can bind the $i$-th target. While in general the matrix may contain
some description of the strength of the interaction, such as the
association constant of the complex, we will focus in this paper on a
simplified description that only differentiates binding from
non-binding molecules, which results in a binary matrix of
target-molecule pairs. This matrix is already sparsely filled with our
current knowledge of protein-ligand interactions, and chemogenomics
attempts to fill the holes. While classical docking or ligand-based
virtual screening strategies focus on each single row independently
from the others in this matrix, i.e., treat each target independently
from each others, the chemogenomics approach is motivated by the
observation that similar molecules can bind similar proteins, and that
information about a known interaction between a ligand and a GPCR
could therefore be a useful hint to predict interaction between
similar molecules and similar GPCRs. This can be of particular
interest when, for example, a particular target has few or no known
ligands, but similar proteins have many: in that case it is tempting
to use the information about the known ligands of similar proteins for
a ligand-based virtual screening of the target of interest. In this
context, we can formally define \textit{in silico} chemogenomics as
the problem of predicting interactions between a molecule and a ligand
(i.e., a hole in the matrix) from the knowledge of all other known
interactions or non-interactions (i.e., the known entries of the
matrix).

Recent
reviews~\citep{Kubinyi2004Chemogenomics,2006Chemical,Klabunde2007Chemogenomic,Rognan2007Chemogenomic}
describe several strategies for \textit{in silico} chemogenomics. A
first class of approaches, called \emph{ligand-based chemogenomics} by
\citet{Rognan2007Chemogenomic}, pool together targets at the level of
families (such as GPCR) or subfamilies (such as purinergic GPCR) and
learn a model for ligands at the level of the family
\citep{Balakin2002Property-based,Klabunde2006ChemogenomicsA}. Other
approaches, termed \emph{target-based chemogenomic} approaches by
\citet{Rognan2007Chemogenomic}, cluster receptors based on ligand
binding site similarity and again pool together known ligands for each
cluster to infer shared ligands \citep{Frimurer2005physicogenetic}.
Finally, a third strategy termed \emph{target-ligand} approach by
\citet{Rognan2007Chemogenomic} attempts to predict ligands for a given
target by leveraging binding information for other targets in a single
step, that is, without first attempting to define a particular set of
similar receptors. This strategy was pioneered by
\citet{Bock2005Virtual} to predict ligands of orphan GPCR. They merged
descriptors of ligands and targets to describe putative
ligand-receptor complexes, and used SVM to discriminate real complexes
from ligand-receptors pairs that do not form complexes.
\cite{Erhan2006Collaborative} followed a similar idea with different
descriptors, and showed in particular that the SVM formulation allows
to generalize the use of vectors of descriptors to the use of positive
definite kernels to describe the chemical and the biological space in
a computationally efficient framework. \citet{Erhan2006Collaborative}
were not able to show, however, significant benefits with respect to
the individual approach that learns a separate classifier for each
GPCR (except in the case of orphan GPCRs, for which their approach
performed better than the baseline random classifier). Recently, in
the context of predicting interactions between peptides and different
alleles of MHC-I molecules, \citet{Jacob2008Efficient} followed a
similar approach and highlighted the importance of choosing adequate
descriptors for small molecules and targets. They obtained
state-of-the-art prediction accuracy for most MHC-I allele, in
particular for those with few known binding peptides.

In this paper we go one step further in this direction and present an
\textit{in silico} chemogenomics approach specifically tailored for
the screening of GPCRs, although the method could in principle be
adapted to other classes of therapeutic targets. We follow the idea of
\citet{Bock2005Virtual} and the algorithmic trick of
\citet{Erhan2006Collaborative}, which allows us to systematically test
a variety of descriptors for both the molecules and the GPCRs. We test
two families of 2D and 3D descriptors to describe molecules, including
a new 3D kernel, and six ways to describe GPCRs, including a
description of their relative positions in current hierarchical
classifications of the superfamily, and information about key residues
likely to be in contact with the ligand.  We test the approach on the
data of the GLIDA database~\citep{Okuno2006GLIDA}, which contains
$34686$ reported interactions between human GPCRs and small molecules,
and observe that the choice of the descriptors has a significant
impact on the accuracy of the models. In particular, the best results
are reached when using the description of GPCRs within the
hierarchical classification of the superfamily, combined with a set of
2D descriptors of small molecules.  This allows us to obtain dramatic
improvements of the prediction accuracy with respect to the individual
learning setting. In an experiment where we simulate the prediction of
ligands for orphan GPCRs, we obtain accuracies of $78.1\%$,
significantly above the $50\%$ baseline accuracy of a random
predictor.

\section{Method}
In this section, we first review the methods proposed by
\citet{Bock2005Virtual,Erhan2006Collaborative} for \textit{in silico}
chemogenomics with SVM, before presenting the particular descriptors
we propose to use for molecules and GPCRs within this framework.

\subsection{\emph{In silico} chemogenomics with machine learning}
We consider the problem of predicting interactions between GPCRs and
small molecules. For this purpose we assume that a list of
target/small molecule pairs
$\left\{(t_{1},m_{1}),\ldots,(t_{n},m_{n})\right\}$, known to interact
or not, is given. Such information is often available as a result of
systematic screening campaigns in the pharmaceutical industry, or on
dedicated databases. Our goal is then to create a model to predict,
for any new candidate pair $(t,m)$, whether the small molecule $m$ is
likely to bind the GPCR $t$.

A general method to create the predictive model is to follow these
four steps:
\begin{enumerate}
\item Choose $n_{tar}$ descriptors to represent each GPCR target $t$
  in the biological space by a $n_{tar}$-dimensional vector
  $\Phi_{tar}(t) = (\Phi_{tar}^1(t) , \ldots ,
  \Phi_{tar}^{n_{tar}}(t))$;
\item In parallel, choose $n_{mol}$ descriptors to represent each
  molecule $m$ in the chemical space by a $n_{mol}$-dimensional vector
  $\Phi_{mol}(m) = (\Phi_{mol}^1(m) , \ldots ,
  \Phi_{mol}^{n_{mol}}(m))$;
\item Derive a vector representation of a candidate target/molecule complex $\Phi_{pair}(t,m)$ from the representations of the target $\Phi_{tar}(t)$ and of the molecule $\Phi_{mol}(m)$;
\item Use a statistical or machine learning method to train a
  classifier able to discriminate between binding and non-binding
  pairs, using the training set of binding and non-binding pairs
  $\left\{\Phi_{pair}(t_{1},m_{1}),\ldots,\Phi_{pair}(t_{n},m_{n})\right\}$
\end{enumerate}
While the first two steps (selection of descriptors) may be specific
to each particular chemogenomics problem, the last two steps define
the particular strategy used for \textit{in silico} chemogenomics. For
example, \cite{Bock2001Predicting,Bock2005Virtual} proposed to
concatenate the vectors $\Phi_{tar}(t)$ and $\Phi_{mol}(m)$ to obtain
a $(n_{tar}+n_{mol})$-dimensional vector representation of the
ligand-target complex $\Phi_{pair}(t,m)$, and to use a SVM as a
machine learning engine. \citet{Erhan2006Collaborative} followed a
slightly different strategy for the third step, by forming descriptors
for the pair $(t,m)$ as \textit{product} of small molecule and target
descriptors. More precisely, given a molecule $m$ described by a
vector $\Phi_{mol}(m)$ and a GPCR $t$ described by a vector
$\Phi_{tar}(t)$, the pair $(t,m)$ is represented by the tensor
product:
\begin{equation}\label{eq:tp}
\Phi_{pair}(t,m) = \Phi_{tar}(t) \otimes \Phi_{mol}(m)\,,
\end{equation}
that is, a $(n_{tar}\times n_{mol})$-dimensional vector whose entries
are products of the form $\Phi_{tar}^{i}(t)\times\Phi_{mol}^{j}(m)$,
for $1\leq i\leq n_{tar}$ and $1\leq j\leq n_{mol}$. A SVM is then
used as an inference engine, to estimate a linear function $f(t,m)$ in
the vector space of target/molecule pairs, that takes positive values
for interacting pairs and negative values for non-interacting ones.

The main motivation for using the tensor product \eqref{eq:tp} is that
it provides a systematic way to encode correlations between small
molecule and target features. For example, in the case of binary
descriptors, the product of two features is $1$ if both the molecule
and the target descriptors are $1$, and zero otherwise, which amounts
to encode the simultaneous presence of particular features of the
molecule and of the target that may be important for the formation of
a complex. A potential issue with this approach, however, is that the
size of the vector representation $n_{tar}\times n_{mol}$ for a pair
may be prohibitively large for practical computation and manipulation.
For example, using a vector of molecular descriptors of size $1024$
for molecules, and representing a protein by the vector of counts of
all $2$-mers of amino-acids in its sequence ($d_{t}=20\times 20 =
400$) results in more than 400k dimensions for the representation of a
pair. As pointed out by \citet{Erhan2006Collaborative}, this
computational obstacle can however be overcome when a SVM is used to
train the linear classifier, thanks to a trick often referred to as
the \textit{kernel trick}. Indeed, a SVM does not necessarily need the
explicit computation of the vectors representing the complexes in the
training set to train a model. What it needs, instead, is the inner
products between these vectors, and a classical property of tensor
products is that the inner product between two tensor products
$\Phi_{pair}(t,m)$ and $\Phi_{pair}(t',m')$ is the product of the
inner product between $\Phi_{tar}(t)$ and $\Phi_{tar}(t')$, on the one
hand, and the inner product between $\Phi_{mol}(m)$ and
$\Phi_{mol}(m')$, on the other hand. More formally, this property can
be written as follows:
\begin{align}\label{eq:inptp}
\nonumber  &\left(\Phi_{tar}(t) \otimes \Phi_{mol}(m)\right)^\top
  \left(\Phi_{tar}(t') \otimes \Phi_{mol}(m')\right)\\
 &= \Phi_{tar}(t)^\top \Phi_{tar}(t')  \times \Phi_{mol}(m)^\top \Phi_{mol}(m') \,,
\end{align}
where $u\top v = u_{1}v_{1} + \ldots + u_{d}v_{d}$ denotes the inner
product between two $d$-dimensional vectors $u$ and $v$. In other
words, the SVM does not need to compute the $n_{tar}\times n_{mol}$
vectors to describe each pair, it only computes the respective inner
products in the target and ligand spaces, before taking the product of
both numbers.

This flexibility to manipulate molecule and target descriptors
separately can moreover be combined with other tricks that sometimes
allow to compute efficiently the inner products in the target and
ligand spaces, respectively. Many such inner products, also called
\textit{kernels}, have been developed recently both in computational
biology~\citep{Schoelkopf2004Kernel} and
chemistry~\citep{Kashima2003Marginalized,Gartner2003graph,Mahe2005Graph},
and can be easily combined within the chemogenomics framework as
follows: if two kernels for molecules and targets are given as:
\begin{equation}\label{eq:kernel}
\begin{split}
  K_{mol}(m,m') & = \Phi_{mol}(m)^\top \Phi_{mol}(m'),\\
  K_{tar}(t,t') &=\Phi_{tar}(t)^\top \Phi_{tar}(t'),
\end{split}
\end{equation}
then we obtain the inner product between tensor products, i.e., the kernel between pairs, by:
\begin{equation}
\label{eq:product}
K\left((t,m),(t',m')\right) = K_{tar}(t,t')\times K_{mol}(m,m').
\end{equation}

In summary, as soon as two vectors of descriptors or kernels $K_{lig}$
and $K_{tar}$ are chosen, we can solve the \emph{in silico}
chemogenomics problem with an SVM using the product kernel
\eqref{eq:product} between pairs. The particular descriptors or
kernels used should ideally encode properties related to the ability
of similar molecules to bind similar targets or ligands respectively.

In the next two subsections, we present different possible choices of
descriptors -- or kernels -- for small molecules and GPCRs,
respectively.

\subsection{Descriptors for small molecules}
The problem of explicitly representing and storing small molecules as
finite-dimensional vectors has a long history in chemoinformatics, and
a multitude of molecular descriptors have been
proposed~\citep{Todeschini2002Handbook}. These descriptors include in
particular physicochemical properties of the molecules, such as its
solubility or logP, descriptors derived from the 2D structure of the
molecule, such as fragment counts or structural fingerprints, or
descriptors extracted from the 3D
structure~\citep{Gasteiger2003Chemoinformatics}. Each classical
fingerprint vector and vector representation of molecules define an
explicit ``chemical space'' in which each molecule is represented by a
finite-dimensional vector, and these vector representations can
obviously be used as such to define kernels between
molecules~\citep{Azencott2007One}.  Alternatively, some authors have
recently proposed some kernels that generalize some of these sets of
descriptors and correspond to inner products between large- or even
infinite-dimensional vectors of descriptors. These descriptors encode,
for example, the counts of an infinite number of walks on the graph
describing the 2D structure of the molecules
\citep{Kashima2004Kernels, Gartner2003graph, Mahe2005Graph}, or
various features extracted from the 3D structures
\citep{Mahe2006Pharmacophore,Azencott2007One}.

In this study we select two existing kernels, encoding respectively 2D
and 3D structural information of the small molecules, and propose a
new 3D kernel:
\begin{itemize}
\item \textit{The 2D Tanimoto kernel}. Our first set of descriptors is
  meant to characterize the 2D structure of the molecules. For a small
  molecule $m$, we define the vector $\Phi_{mol}(m)$ as the binary
  vector whose bits indicate the presence or absence of all linear
  graph of length $u$ or less as subgraphs of the 2D structure of $l$.
  We chose $u=8$ in our experiment, i.e., characterize the molecules
  by the occurrences of linear subgraphs of length $8$ or less, a
  value previously observed to give good results in several virtual
  screening tasks~\citep{Mahe2005Graph}. Moreover, instead of directly
  taking the inner product between vectors as in \eqref{eq:kernel}, we
  use the Tanimoto kernel:
\begin{equation}
\label{eq:tanimoto}
K_{ligand}(l,l') = \frac{\Phi_{lig}(l)^\top\Phi_{lig}(c')}{\Phi_{lig}(l)^\top\Phi_{lig}(l)+\Phi_{lig}(l')^\top\Phi_{lig}(l')-\Phi_{lig}(l)^\top\Phi_{lig}(l')}\,,
\end{equation}
which was proven to be a valid inner product
by~\cite{Ralaivola2005Graph}, giving very competitive results on a
variety of QSAR or toxicity prediction experiments.
\item \emph{3D pharmacophore kernel} While 2D structures are known to
  be very competitive in ligand-based virtual screening
  \citep{Azencott2007One}, we reasoned that some specific 3D
  conformations of a few atoms or functional groups may be responsible
  for the interaction with the target. Thus, we decided to test
  descriptors representing the presence of potential 3-point
  pharmacophores. For this, we used the 3D pharmacophore kernel
  proposed by~\cite{Mahe2006Pharmacophore}, that generalizes 3D
  pharmacophore fingerprint descriptors. This approach implies the
  choice of a 3D conformer for each molecule. In absence of sufficient
  data available for bound ligands in GPCR structures, we chose to
  build a 3D version of the ligand base in which molecules are
  represented in an estimated minimum energy conformation. For each of
  the $2446$ retained ligands, $25$ conformers were generated with the
  Omega program (OpenEye Scientific Software) using standard
  parameters, except for a $1\AA$ RMSD clustering of the conformers,
  instead of the $0.8$ default value. A 3D ligand base was generated
  by keeping the conformer of lowest energy for each ligand. Partial
  charges were calculated for all atoms using the molcharge program
  (OpenEye Scientific Software) with standard parameters. This ligand
  base was then used to calculate a 3D pharmacophore kernel for
  molecules~\citep{Mahe2006Pharmacophore}.
\end{itemize}

We used the freely and publicly available
\emph{ChemCPP}\footnote{Available at
  \url{http://chemcpp.sourceforge.net}.} software to compute the 2D
and 3D pharmacophore kernel.

\subsection{Descriptors for GPCRs}
\begin{figure}[ht]
  \centering
  \includegraphics[width=.6\linewidth]{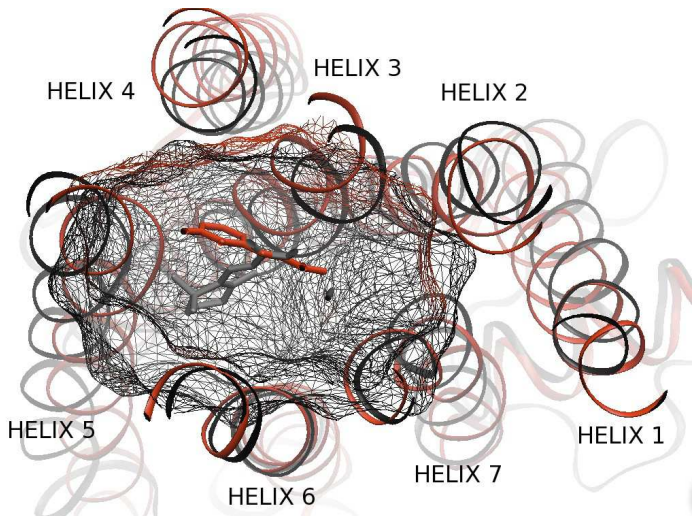}
  \caption{Representation of the binding pocket of
    $\beta_2$-adrenergic receptor (in red) and bovine Rhodopsin (in
    black) viewed from the extracellular surface. On the center of the
    pocket, 3-(isopropylamino)propan-2-ol and cis-retinal have been
    represented to show the size and the position of the pocket around
    each ligand. Figure drawn with VMD~\citep{Humphrey1996VMD:}.}
\label{fig:pocket}
\end{figure}

\begin{figure}[ht]
  \centering
  \includegraphics[width=.6\linewidth]{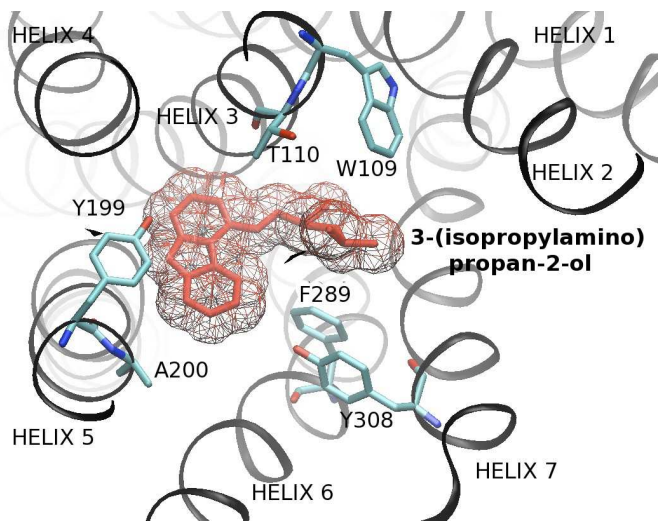}
  \caption{3-(isopropylamino)propan-2-ol and the protein environment
    of $\beta_2$-adrenergic receptor as viewed from the extracellular
    surface. Amino acid side chains are represented for $6$ of the
    $31$ residues (in cyan, blue and red) of the binding pocket motif.
    Transmembranes helix and 3-(isopropylamino)propan-2-ol are colored
    in black and red respectively. Figure drawn with
    VMD~\citep{Humphrey1996VMD:}.}
\label{fig:ligand}
\end{figure}

SVM and kernel methods are also widely used in
bioinformatics~\citep{Schoelkopf2004Kernel}, and a variety of
approaches have been proposed to design kernels between proteins,
ranging from kernels based on the amino-acid sequence of a
protein~\citep{Jaakkola2000Discriminative,Leslie2002spectrum,
  Tsuda2002Marginalized,Leslie2004Mismatch,
  Vert2004Local,Kuang2005Profile-based,Cuturi2005context-tree} to
kernels based on the 3D structures of
proteins~\citep{Dobson2005Predicting,Borgwardt2005Protein,Qiu2007structural}
or on the pattern of occurrences of proteins in multiple sequenced
genomes~\citep{Vert2002tree}. These kernels have been used in
conjunction with SVM or other kernel methods for various tasks related
to structural or functional classification of proteins. While any of
these kernels can theoretically be used as a GPCR kernel
in~\eqref{eq:product}, we investigate in this paper a restricted list
of specific kernels described below, aimed at illustrating the
flexibility of our framework and test various hypothesis.
\begin{itemize}
\item The \emph{Dirac} kernel between two targets $t,t'$ is:
\begin{equation}
  K_{Dirac}(t,t') = \begin{cases}
    1 &\text{ if }t=t'\,,\\
    0&\text{ otherwise.}
\end{cases}
\end{equation}
This basic kernel simply represents different targets as orthonormal
vectors. From (\ref{eq:product}) we see that orthogonality between two
proteins $t$ and $t'$ implies orthogonality between all pairs $(l,t)$
and $(l',t')$ for any two small molecules $c$ and $c'$. This means
that a linear classifier for pairs $(l,t)$ with this kernel decomposes
as a set of independent linear classifiers for interactions between
molecules and each target protein, which are trained without sharing
any information of known ligands between different targets. In other
words, using Dirac kernel for proteins amounts to performing classical
learning independently for each target, which is our baseline
approach.
\item The \emph{multitask} kernel between two targets $t,t'$ is
  defined as:
  \begin{displaymath}
        K_{multitask}(t,t') = 1+K_{Dirac}(t,t')\,.
  \end{displaymath}
  This kernel, originally proposed in the context of multitask
  learning \cite{Evgeniou2005Learning}, removes the orthogonality of
  different proteins to allow sharing of information. As explained in
  \cite{Evgeniou2005Learning}, plugging $K_{multitask}$ in
  \eqref{eq:product} amounts to decomposing the linear function used
  to predict interactions as a sum of a linear function common to all
  GPCRs and of a linear function specific to each GPCR:
  \begin{displaymath}
  f(l,t) = w^\top \Phi(l,t) = w_{general}^\top \Phi_{lig}(l) + w_{t}^\top\Phi_{lig}(l)\,.
  \end{displaymath}
  A consequence is that only data related to the the target $t$ are
  used to estimate the specific vector $w_{t}$, while all data are
  used to estimate the common vector $w_{general}$. In our framework
  this classifier is therefore the combination of a target-specific
  part accounting for target-specific properties of the ligands and a
  global part accounting for general properties of the ligands across
  the targets.  The latter term allows to share information during the
  learning process, while the former ensures that specificities of the
  ligands for each target are not lost.  

\item The \textit{hierarchy} kernel.  Alternatively we could propose a
  new kernel aimed at encoding the similarity of proteins with respect
  to the ligands they bind.  In the GLIDA database indeed, GPCRs are
  grouped into $4$ classes based on sequence homology and functional
  similarity: the \emph{rhodopsin} family (class A), the
  \emph{secretin} family (class B), the \emph{metabotropic} family
  (class C) and some smaller classes containing other GPCRs. The GLIDA
  database further subdivides each class of targets by type of
  ligands, for example amine or peptide receptors or more specific
  families of ligands. This also defines a natural hierarchy that can
  be used to compare GPCRs.

  The hierarchy kernel between two GPCRs was therefore defined as the
  number of common ancestors in the corresponding hierarchy plus one,
  that is,
  \begin{displaymath}
    K_{hierarchy}(t,t') = \langle\Phi_h(t),\Phi_h(t')\rangle,
  \end{displaymath}
  where $\Phi_h(t)$ contains as many features as there are nodes in
  the hierarchy, each being set to $1$ if the corresponding node is
  part of $t$'s hierarchy and $0$ otherwise, plus one feature
  constantly set to one that accounts for the "plus one" term of the
  kernel.
\item The \emph{binding pocket} kernel. Because the protein-ligand
  recognition process occurs in 3D space in a pocket involving a
  limited number of residues, we tried to describe the GPCR space
  using a representation of this pocket. The difficulty resides in the
  fact that although the GPCR sequences are known, the residues
  forming this pocket and its precise geometry are \emph{a priori}
  unknown. However, the two available X-Ray structures, together with
  mutagenesis data showed that the binding pockets are situated in a
  similar region for all GPCRs~\citep{Kratochwil2005automated}. In
  order to identify residues potentially involved in the binding
  pocket of GPCRs of unknown structure studied in this work, we
  proceeded in several steps. (a) The two known structures (PDB
  entries 1U19 and 2RH1) were superimposed using the STAMP
  algorithm~\citep{Russell1992Multiple}. In the superimposed
  structures, the retinal and 3-(isopropylamino)propan- 2-ol ligands
  are very close, which is in agreement with global conservation of
  binding pockets, as shown on Figure~\ref{fig:pocket}.  (b) The
  structural alignment of bovine rhodopsin and of human
  $\beta_2$-adrenergic receptor was used to generate a sequence
  alignment of these two proteins.  (c) For both structures, in order
  to identify residues potentially involved in stabilizing
  interactions with the ligand (residues of the pocket), we selected
  residues that presented at least one atom situated at less than $6
  \AA$ from at least one atom of the ligand. Figure~\ref{fig:ligand}
  shows that these two pockets clearly overlap, as expected. (d)
  Residues of the two pockets (as defined in (c)) were labeled in this
  structural sequence alignment. These residues were found to form
  small sequence clusters that were in correspondence in this
  alignment. These clusters were situated mainly in the apical region
  of transmembrane segments and included a few extracellular residues.
  (e) All studied GPCR sequences, including bovine rhodopsin and of
  human $\beta_2$-adrenergic receptor were aligned using
  CLUSTALW~\citep{Chenna2003Multiple} with Blosum
  matrices~\citep{Henikoff1992Amino}.  For each protein, residues in
  correspondence with a residue of the binding pocket (as defined
  above) of either bovine rhodopsin or human $\beta_2$-adrenergic
  receptor were retained.  This lead to a different number of residues
  per protein, because of sequence variability. For example, in
  extracellular regions, some residues from bovine rhodopsin or human
  $\beta_2$-adrenergic receptor had a corresponding residue in some
  sequences but not in others. In order to provide a homogeneous
  description of all GPCRs, in the list of residues initially retained
  for each protein, only residues situated at positions conserved in
  almost all GPCRs were kept. (f) Each protein was then represented by
  a vector whose elements corresponded to a potential conserved
  pocket. This description, although appearing as a linear vector
  filled with amino acid residues, implicitly codes for a 3D
  information on the receptor pocket, as illustrated on
  Figure~\ref{fig:ligand}. These vectors were then used to build a
  kernel that allows comparison of binding pockets.  The classical way
  to represent motifs of constant length as fixed length vectors is to
  encode the letter at each position by a $20$-dimensional binary
  vector indicating which amino acid is present, resulting in a
  $180$-dimensional vector representations. In terms of kernel, the
  inner product between two binding pocket motifs in this
  representation is simply the number of letters they have in common
  at the same positions:
  \begin{displaymath}
    K_{pb}(x,x') = \sum_{i=1}^{l}\delta(x[i],x'[i]),
  \end{displaymath}
  where $l$ is the length of the binding pocket motifs ($31$ in our
  case), $x[i]$ is the $i$-th residue in x and $\delta(x[i],x'[i])$ is
  $1$ if $x[i] = x'[i]$, $0$ otherwise. This is the baseline pocket
  binding kernel.  Alternatively, using a polynomial kernel of degree
  $p$ over the baseline kernel is equivalent, in terms of feature
  space, to encoding $p$-order interactions between amino acids at
  different positions. In order to assess the relevance of such
  non-linear extensions we tested this polynomial pocket binding
  kernel,
  \begin{displaymath}
    K_{ppb}(x,x') = \left(K_{pb}(x,x')+1\right)^p.
  \end{displaymath}
  We only used a degree $p=2$, although a more careful choice of this
  parameter could further improve the performances.
\item The \emph{binding pocket hierarchy} kernel. Because of the link
  between binding pockets and ligand recognition, we also defined a
  new hierarchy based on the sequence alignment of the binding pocket
  amino acid vectors without gaps. To do this, we used a PAM matrix
  with high values of gap insertion and extension to compare each
  couple of GPCR vectors. The obtained scores were used in UPGMA
  (Unweighted Pair Group Method with Arithmetic mean) to determine a
  binding pocket similarity based hierarchy.  We obtained a tree
  comparable to phylogenetic trees, and that happens to be share many
  substructures with the GLIDA hierarchy.
  \end{itemize}

\section{Data}
We used the GLIDA GPCR-ligand database~\citep{Okuno2006GLIDA} which
includes $22964$ known ligands for and $3738$ GPCRs from human, rat
and mouse. The ligand base contains highly diverse molecules, from
ions and very small molecules up to peptides. In order to eliminate
unwanted molecules such as inorganic compounds and molecules with
unsuitable molecular weights, we filtered the GLIDA ligand base using
the filter program (OpenEye Scientific Software) with standard
parameters. The most important filtering feature here was to keep
molecules of molecular weights ranging from $150$ Da to $450$ Da.
Overall, the GLIDA ligand base was filtered in order to retain
molecules that had the physico- chemical characteristics of drugs.
This filter retained $2688$ molecules. Because the GLIDA ligand base
contains a few duplicates, we eliminated these redundancies, which
lead to $2446$ different molecules, available under a 2D description
files and giving $4051$ interactions with the human GPCRs. Elimination
of duplicates present in the GLIDA base was important here because it
could have lead to overfitting in the learning step. For each positive
interaction given by this restricted set, we generated a negative
interaction involving the same receptor and one of the ligands that
was in the database and was not indicated as one of its ligands. This
probably generated some false negative points in our benchmark, and it
would be interesting to use experimentally tested negative
interactions. We loaded the sequences of all GPCRs that are able to
bind any of these ligands, which lead to 80 sequences, all
corresponding to human GPCRs.  In the GLIDA database, GPCRs are
classified in a hierarchy (as mentioned above) which was also loaded
for use in the hierarchy kernel.

\section{Results}
We ran two different sets of experiments on this dataset in order to
illustrate two important points. In a first set of experiments, for
each GPCR, we 5-folded the data available, i.e. the line of the
interaction matrix corresponding to this GPCR. The classifier was
trained with four folds and the whole data from the other GPCRs,
\emph{i.e.}, all other lines of the interaction matrix. The prediction
accuracy for the GPCR under study was then tested on the remaining
fold. The goal of these first experiments was to evaluate if using
data from other GPCRs improved the prediction accuracy for a given
GPCR. In a second set of experiments, for each GPCR we trained a
classifier on the whole data from the other GPCRs, and tested on the
data of the considered GPCR.  The goal was to assess how efficient our
chemogenomics approach would be to predict the ligands of orphan
GPCRs. In both experiments, the $C$ parameter of the SVM was selected
by internal cross validation on the training set among $2^i,
i\in\{-8,-7,\ldots,5,6\}$.

For the first experiment, since learning an SVM with only one training
point does not really make sense and can lead to "anti-learning" less
than $0.5$ performances, we set all results $r$ involving the Dirac
GPCR kernel on GPCRs with only $1$ known ligand to $\max(r,0.5)$.
This is to avoid any artefactual penalization of the Dirac approach
and make sure we measure the actual improvement brought by sharing
information across GPCRs.

\begin{table}[!t]
  \centering
  \begin{tabular}{l c c}\hline
      $K_{tar}\backslash K_{lig}$ & 2D Tanimoto & 3D pharmacophore \\\hline
      Dirac  &  $86.2\pm1.9$ & $84.4\pm2.0$\\
      multitask &   $88.8\pm1.9$ & $85.0\pm2.3$\\
      hierarchy &   $93.1\pm1.3$ & $88.5\pm2.0$\\
      binding pocket  &  $90.3\pm1.9$ & $87.1\pm2.3$\\
      poly binding pocket &   $92.1\pm1.5$ & $87.4\pm2.2$\\
      binding pocket hierarchy & $93.0\pm1.4$ & $90.0\pm2.1$\\\hline
  \end{tabular}
  \caption{Prediction accuracy for the first experiment with
    various ligand and target kernels.}
  \label{tab:exp1}
\end{table}

\begin{figure*}[ht]
  \centering
  \begin{minipage}[b]{.23\linewidth}
  \centering
  \includegraphics[width=\linewidth]{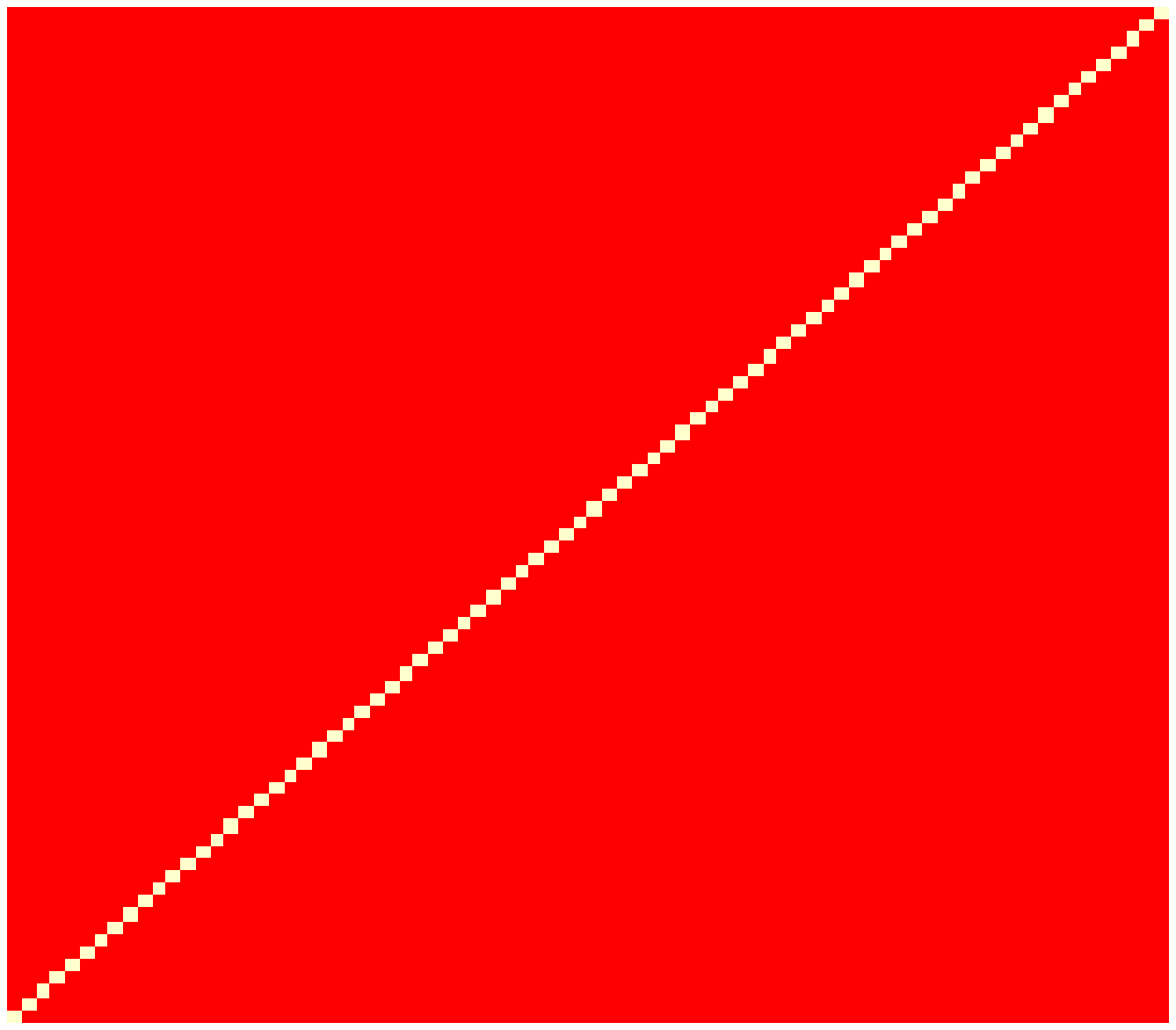}
  \end{minipage}
  \hspace{0.2cm}
  \begin{minipage}[b]{.23\linewidth}
  \centering
  \includegraphics[width=\linewidth]{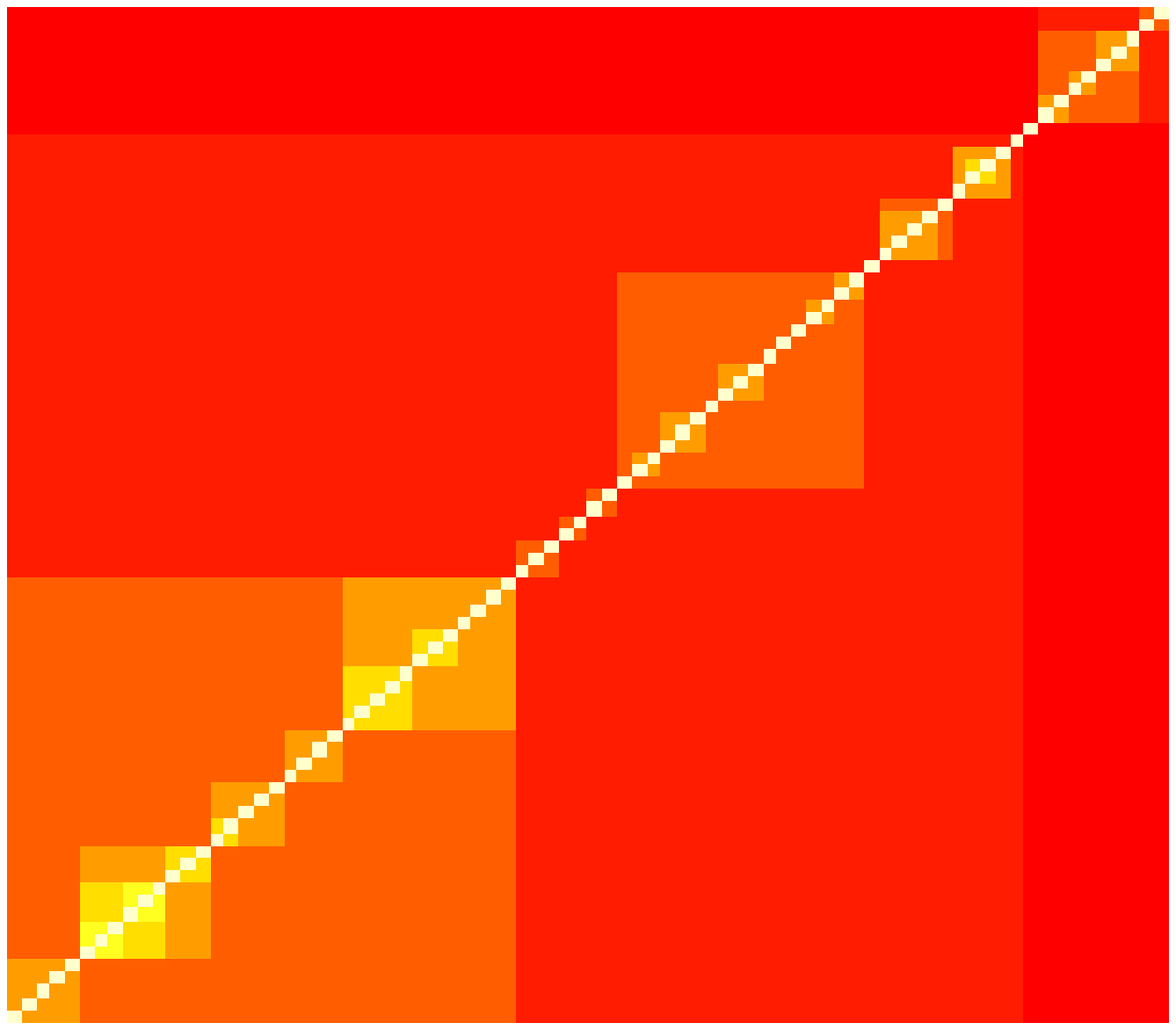}
  \end{minipage}
  \hspace{0.2cm}
  \begin{minipage}[b]{.23\linewidth}
  \centering
  \includegraphics[width=\linewidth]{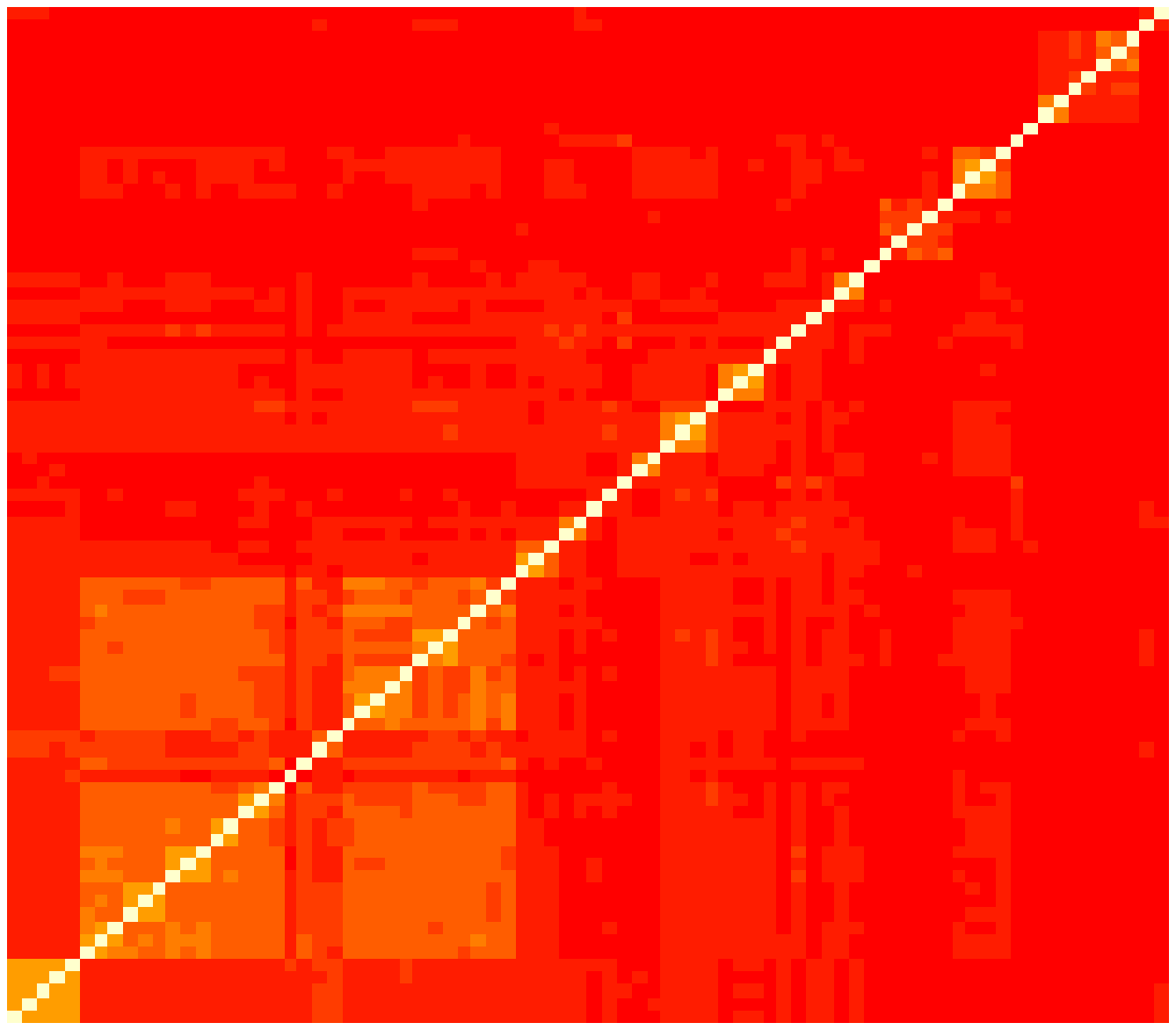}
  \end{minipage}
  \begin{minipage}[b]{.23\linewidth}
  \centering
  \includegraphics[width=\linewidth]{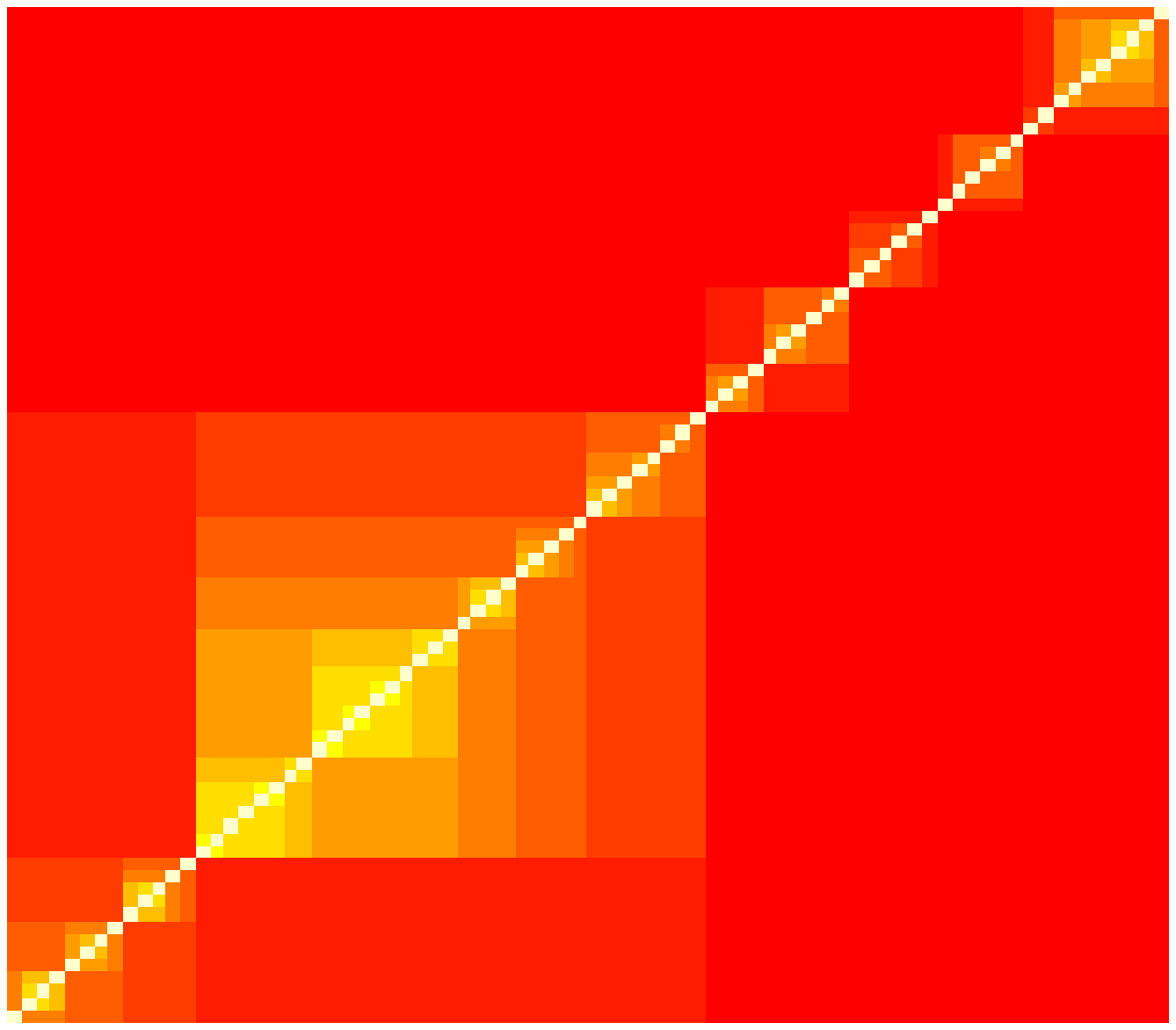}
  \end{minipage}
  \caption{GPCR kernel Gram matrices ($K_{tar}$) for the GLIDA GPCR
    data with multitask, hierarchy, binding pocket and binding pocket
    hierarchy kernels.}
  \label{fig:kernels}
\end{figure*}

Table~\ref{tab:exp1} shows the results of the first experiments with
all the ligand and GPCR kernel combinations. For all the ligand
kernels, one observes an improvement between the individual approach
(Dirac GPCR kernel, $86.4\%$) and the baseline multitask approach
(multitask GPCR kernel, $88.4\%$). The latter kernel is merely
modeling the fact that each GPCR is uniformly similar to all other
GPCRs, and twice more similar to itself. It does not use any prior
information on the GPCRs, and yet, using it improves the global
performance with respect to individual learning. Using more
informative GPCR kernels further improves, sometimes considerably, the
prediction accuracy. In particular, the hierarchy kernels add more
than $4.5\%$ of precision with respect to naive multitask approach.
All the other informative GPCR kernels also improve the performance.
The polynomial binding pocket kernel and the pocket binding hierarchy
kernels are almost as efficient as the hierarchy kernel, which is an
interesting result. Indeed, one could fear that using the hierarchy
kernel, for the construction of which some knowledge of the ligands
may have been used, could have introduced bias in the results.  Such
bias is certainly absent in the binding pocket kernel. The fact that
the same performance can be reached with kernels based on the mere
sequence of GPCRs' pockets is therefore an important result.
Figure~\ref{fig:kernels} shows four of the GPCR kernels. The baseline
multitask is shown as a comparison.  Interestingly, many of the
subgroups defined in the hierarchy can be found in the binding pocket
kernel, that is, they are retrieved from the simple information of the
binding pocket sequence. This phenomenon is even more visible for the
binding pocket hierarchy kernel that is based on the hierarchy built
from the binding pocket alignment scores.

\begin{figure}[ht]
  \centering
  \includegraphics[width=.6\linewidth]{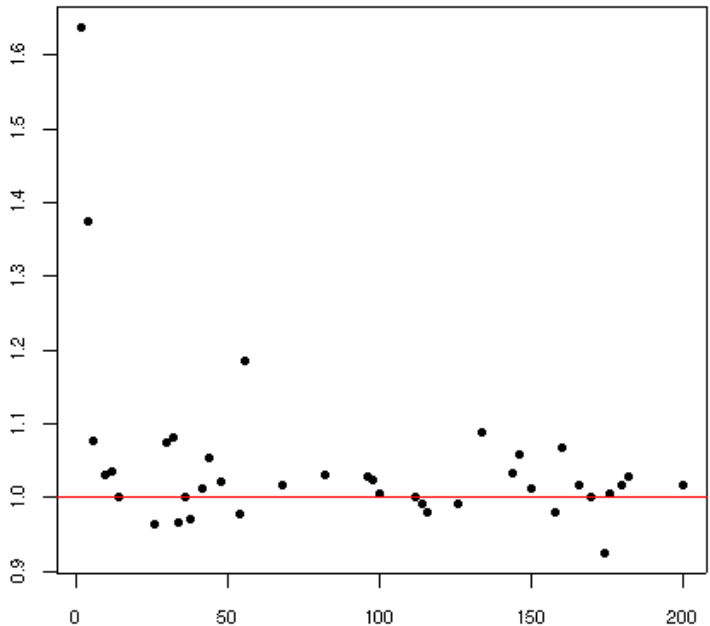}
  \caption{Improvement (as a performance ratio) of the hierarchy GPCR
    kernel against the Dirac GPCR kernel as a function of the number
    of training samples available. Restricted to $[2-200]$ samples
    for the sake of readability.}
\label{fig:dvsh}
\end{figure}

The 3D kernel for the ligands, on the other hand, did not perform as
well as the 2D kernel. This can be either explained by the fact the
the pharmacophore kernel is not suited to this problem, or by the fact
that choosing the conformer of the ligand is not a trivial task. This
point is discussed below.

\begin{figure}[ht]
  \centering
  \includegraphics[width=.6\linewidth]{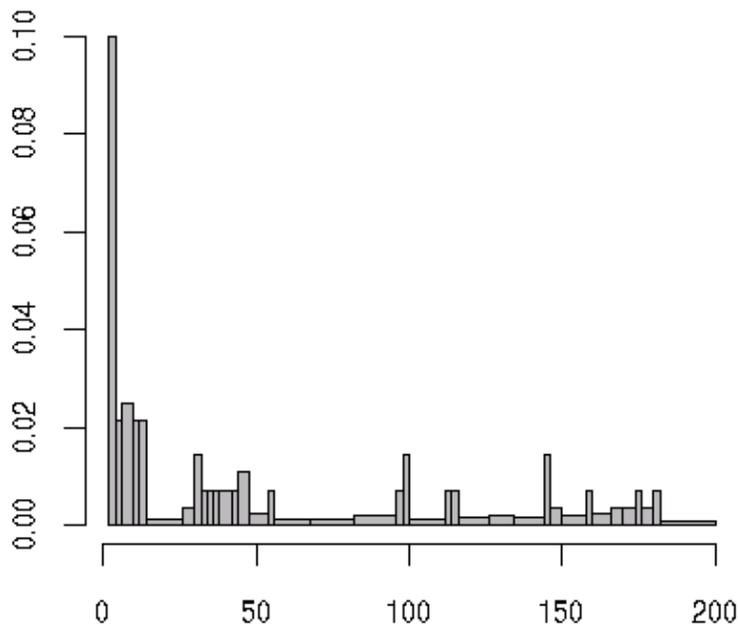}
  \caption{Distribution of the number of training points for a GPCR.
    Restricted to $[2-200]$ samples for the sake of readability.}
\label{fig:dist}
\end{figure}

Figure~\ref{fig:dvsh} illustrates how the improvement brought by the
chemogenomics approach varies with the number of available training
points. As one could have expected, the strongest improvement is
observed for the GPCRs with few (less than $20$) training points
(\emph{i.e.}, less than $10$ known ligands since for each known ligand
an artificial non-ligand was generated). When more training points
become available, the improvement is less important, and sharing the
information across the GPCRs can even degrade the performances. This
is an important point, first because, as showed on
Figure~\ref{fig:dist}, many GPCRs have few known ligands (in
particular, $11$ of them have only two training points), and second
because it shows that when enough training points are available,
individual learning will probably perform as well as or better than
our chemogenomics approach.

Our second experiment intends to assess how our chemogenomics approach
can perform when predicting ligands for orphan GPCRs, \emph{i.e.},
with no training data available for the GPCR of interest.
Table~\ref{tab:exp2} shows that in this setting, individual learning
performs random prediction. Naive multitask approach does not improve
much the performance, but informative kernels such as hierarchical and
binding pocket kernels achieve $77.4\%$ and $78.1\%$ of precision
respectively, that is, almost $30\%$ better than the random approach
one would get when no data is available. Here again, the fact that the
binding pocket kernel that only uses the sequence of the receptor
pocket performs as well as the hierarchy-based kernel is encouraging.
It suggests that given a receptor for which nothing is known except
its sequence, it is possible to make reasonable ligand predictions.

\begin{table}[!t]
  \centering
  \begin{tabular}{l c c}\hline
      $K_{tar}\backslash K_{lig}$ & 2D Tanimoto & 3D pharmacophore \\\hline
      Dirac & $50.0\pm0.0$ & $50.0\pm0.0$\\
      multitask & $56.8\pm2.5$ & $58.2\pm2.2$\\
      hierarchy & $77.4\pm2.4$ & $76.2\pm2.2$\\
      binding pocket & $78.1\pm2.3$ & $76.6\pm2.2$\\
      poly binding pocket & $76.4\pm2.4$ & $74.9\pm2.3$\\
      binding pocket hierarchy & $75.5\pm2.4$ & $76.5\pm2.2$\\\hline
  \end{tabular}
  \caption{Prediction accuracy for the second experiment with
    various ligand and target kernels.}
  \label{tab:exp2}
\end{table}

\section{Discussion}

We showed how sharing information across the GPCRs by considering a
chemogenomics space of the GPCR-ligand interaction pairs could improve
the prediction performances. In addition, we showed that using such a
representation, it was possible to make reasonable predictions even
when no ligand was known for a given GPCR, that is, to predict ligands
for orphan GPCRs. Our approach is simply to apply well known machine
learning methods in the constructed chemogenomics space. We used a
systematic way to build such a space by combining a given
representation of the ligands with a given representation of the GPCRs
into a binding-prediction-oriented GPCR-ligand couple representation.
This allows to use any ligand or GPCR descriptor or kernel existing in
the chemoinformatics or bioinformatics literature, or new ones
containing other prior information as we tried to propose in this
paper. Our experiments showed that the choice of the descriptors was
crucial for the prediction, and more sophisticated features for either
the ligands or the GPCRs could probably further improve the
performances.

In all experiments, 3D pharmacophore kernels did not reach the
performance of 2D kernels for the ligands. This is apparently in
contradiction we the idea that protein-ligand interaction is a process
occurring in the 3D space, and that introduction of 3D information
should increase the performance. Different explanations can be
proposed. The choice of the low energy conformer was guided by the
following idea. Because only two ligand conformations bound to GPCR
receptors are available, it was not possible to derive any general
information that could be used to choose a potential bioactive 3D
conformer for each molecule of the ligand base. In this context, the
only possible reasonable assumption was that, while interaction with
the receptor will certainly perturb the conformational energy surface
of a flexible ligand, high affinity would be observed for ligands that
bind in a conformation that is not exceptionally different from a
local free state energy minimum~\citep{Bostroem2001Reproducing}.
Although there exists a large number of methods for exploring the
conformation space of a molecule, we used the Omega program that
performs rapid systematic conformer search, because it has been showed
to present good performances for retrieving bioactive
conformations~\citep{Bostroem2003Assessing}. However, the set of
parameters used to run Omega in this study (because of calculation
time limitations) may not have allowed to reach a local energy
minimum: generating a larger number of conformers, with a smaller RMSD
clustering value may have helped to find better energy minima, and
this could be further evaluated.  Moreover, some studies report that
the bioactive conformation of a molecule can differ from the minimum
energy conformation, and that significant strain energies can indeed
be found for molecules in complex with
proteins~\citep{Perola2004Conformational}. We cannot rule out the
possibility that this is the case for GPCR ligands. In the future,
resolution of additional 3D structures in this family will help to
clarify this point. One possible improvement of the method could be to
use homology models for the GPCRs, dock the ligand base in the modeled
binding pockets, and build a 3D ligand base using, for each molecule,
the conformer associated to the best docking solution. In other
families of proteins, enzymes for example, where many structures are
available and can be used to define bioactive conformers, the 3D
pharmacophore kernel is expected to improve performance, as observed
in a previous pure ligand-based study involving ligands in a given
series, for which the bioactive conformation can be inferred from a
known 3D structure~\citep{Mahe2006Pharmacophore}.

Various evidence suggest that, within a common global architecture, a
generic binding pocket mainly involving transmembrane regions hosts
agonists, antagonists and allosteric modulators. In order to identify
this pocket automatically, other studies report the use of sequence
alignment and the prediction of transmembrane helices.
\cite{Kratochwil2005automated} detected hypervariable positions in
transmembrane helices for identification of residues forming the
binding pocket. The underlying idea was that conserved residues were
probably important for structure stabilization, while variable
positions were involved in ligand binding, in order to accommodate the
wide spectrum of molecules that are GPCR substrates. Using this
method, they proposed potential binding pockets for GPCRs, and found
that the corresponding residues were frequently in the GRAP mutant
database for GPCRs~\citep{Kristiansen1996database}. Interestingly,
these authors pointed that residues corresponding to these
hypervariable positions were found within a distance of $6\AA$ from
retinal in the rhodopsin X-Ray structure. Therefore, although we used
a different method to automatically extract binding pocket residues in
the GPCR families, our results are in good agreement with this study.

Interesting developments of this method could include application to
quantitative prediction of the binding affinities, that would be
straightforward using regression algorithms in the same chemogenomics
space. Another possibility is application to other important drug
target families, like enzymes or ion
channels~\citep{Hopkins2002druggable}, for which most of the
descriptors used for the GPCRs in this paper could directly be used,
and other, more specific ones could be designed. From a methodological
point of view, many recent developments in multitask
learning~\citep{Vert2006Low-rank,Argyriou2007Multi-task,Bonilla2008Multi-task}
could be applied to generalize this chemogenomics approach using, for
example, other regularization methods.

\begin{sidewaystable}[!t]
   \begin{tabular}{l*{15}{ c}}\hline
      positions on $\beta_2$-adrenergic receptor & $82$ & $\mathbf{109}$ & $\mathbf{110}$ & $113$ & $114$ & $115$ & $116$ & $117$ & $118$ &
      $121$ & $175$ & $183$ & $195$ & $\mathbf{199}$ & $\mathbf{200}$ \\\hline
      $\beta_2$-adrenergic receptor & M & \textbf{W} & \textbf{T} & D & V & L & C & V & T & I & R & N & T & \textbf{Y} & \textbf{A}\\
      5-hydroxytryptamine 5A receptor & V & \textbf{W} & \textbf{I} & D & V & L & C & C & T & I & I & E & S & \textbf{Y} & \textbf{A}\\
      Adenosine A2b receptor & V & \textbf{L} & \textbf{A} & V & L & V & L & T & Q & I & I & K & K & \textbf{M} & \textbf{V}\\
      Gamma-aminobutyric acid type B receptor & E & \textbf{D} & \textbf{E} & E & A & V & E & G & H & T & L & G & S & \textbf{F} & \textbf{D}\\
      Relaxin 3 receptor 2 & L & \textbf{V} & \textbf{L} & T & V & L & N & V & Y & I & V & G & L & \textbf{Y} & \textbf{Q}\\\hline
      \\
      \end{tabular}
      \begin{tabular}{l*{16}{ c}}\hline
      positions on $\beta_2$-adrenergic receptor & $203$ & $204$ & $207$
      & $208$ & $212$ & $282$ & $286$ & $\mathbf{289}$ & $290$ & $293$ & $\mathbf{308}$ & $311$ & $312$ & $313$ & $315$ & $316$\\\hline
      $\beta_2$-adrenergic receptor & S & S & S & F & L & F & W & \textbf{F} & F & N & \textbf{Y} & L & N & W & G & Y\\
      5-hydroxytryptamine 5A receptor  & S & T & A & F & L & F & W & \textbf{F} & F & E & \textbf{K} & F & L & W & G & Y\\
      Adenosine A2b receptor & N & F & C & V & L & F & W & \textbf{V} & H & N & \textbf{M} & A & I & L & S & H\\
      Gamma-aminobutyric acid type B receptor & G & S & A & W & E & F & L & \textbf{Y} & H & R & \textbf{L} & T & V & G & L & V\\
      Relaxin 3 receptor 2 & R & V & A & F & L & F & W & \textbf{N} & H & T & \textbf{F} & T & T & C & A & H\\\hline
      \end{tabular}
  \caption{Residues of 5-hydroxytryptamine 5A receptor, Adenosine A2b
    receptor, Gamma-aminobutyric acid type B receptor and Relaxin 3
    receptor 2 (shown as examples) aligned with $\beta_2$-adrenergic
    receptor binding site amino acids. the binding pocket motif of
    $\beta_2$-adrenergic receptor has been used as reference to determine
    residues involved in the formation of the binding site of the $79$
    other GPCRs. Bold columns correspond to the residues shown on
    Figure~\ref{fig:ligand}.}
  \label{tab:residues}
\end{sidewaystable}


\begin{thebibliography}{}

\bibitem[Argyriou {\em et~al.}(2007)Argyriou, Evgeniou, and
  Pontil]{Argyriou2007Multi-task}
Argyriou, A., Evgeniou, T., and Pontil, M. (2007).
\newblock Multi-task feature learning.
\newblock In B.~Sch\"{o}lkopf, J.~Platt, and T.~Hoffman, editors, {\em Adv.
  Neural. Inform. Process Syst. 19\/}, pages 41--48, Cambridge, MA. MIT Press.

\bibitem[Azencott {\em et~al.}(2007)Azencott, Ksikes, Swamidass, Chen,
  Ralaivola, and Baldi]{Azencott2007One}
Azencott, C.-A., Ksikes, A., Swamidass, S.~J., Chen, J.~H., Ralaivola, L., and
  Baldi, P. (2007).
\newblock One- to four-dimensional kernels for virtual screening and the
  prediction of physical, chemical, and biological properties.
\newblock {\em J Chem Inf Model\/}, {\bf 47}(3), 965--974.

\bibitem[Balakin {\em et~al.}(2002)Balakin, Tkachenko, Lang, Okun, Ivashchenko,
  and Savchuk]{Balakin2002Property-based}
Balakin, K.~V., Tkachenko, S.~E., Lang, S.~A., Okun, I., Ivashchenko, A.~A.,
  and Savchuk, N.~P. (2002).
\newblock Property-based design of gpcr-targeted library.
\newblock {\em J Chem Inf Comput Sci\/}, {\bf 42}(6), 1332--1342.

\bibitem[Bock and Gough(2001)Bock and Gough]{Bock2001Predicting}
Bock, J.~R. and Gough, D.~A. (2001).
\newblock Predicting protein-protein interactions from primary structure.
\newblock {\em Bioinformatics\/}, {\bf 17}(5), 455--460.

\bibitem[Bock and Gough(2005)Bock and Gough]{Bock2005Virtual}
Bock, J.~R. and Gough, D.~A. (2005).
\newblock Virtual screen for ligands of orphan g protein-coupled receptors.
\newblock {\em J Chem Inf Model\/}, {\bf 45}(5), 1402--1414.

\bibitem[Bockaert and Pin(1999)Bockaert and Pin]{Bockaert1999Molecular}
Bockaert, J. and Pin, J.~P. (1999).
\newblock Molecular tinkering of g protein-coupled receptors: an evolutionary
  success.
\newblock {\em EMBO J\/}, {\bf 18}(7), 1723--1729.

\bibitem[Bonilla {\em et~al.}(2008)Bonilla, Chai, and
  Williams]{Bonilla2008Multi-task}
Bonilla, E., Chai, K.~M., and Williams, C. (2008).
\newblock Multi-task gaussian process prediction.
\newblock In J.~Platt, D.~Koller, Y.~Singer, and S.~Roweis, editors, {\em
  Advances in Neural Information Processing Systems 20\/}. MIT Press,
  Cambridge, MA.

\bibitem[Borgwardt {\em et~al.}(2005)Borgwardt, Ong, Sch{\"o}nauer,
  Vishwanathan, Smola, and Kriegel]{Borgwardt2005Protein}
Borgwardt, K., Ong, C., Sch{\"o}nauer, S., Vishwanathan, S., Smola, A., and
  Kriegel, H.-P. (2005).
\newblock Protein function prediction via graph kernels.
\newblock {\em Bioinformatics\/}, {\bf 21}(Suppl. 1), i47--i56.

\bibitem[Bostr\"om(2001)Bostr\"om]{Bostroem2001Reproducing}
Bostr\"om, J. (2001).
\newblock Reproducing the conformations of protein-bound ligands: a critical
  evaluation of several popular conformational searching tools.
\newblock {\em J Comput Aided Mol Des\/}, {\bf 15}(12), 1137--1152.

\bibitem[Bostr\"om {\em et~al.}(2003)Bostr\"om, Greenwood, and
  Gottfries]{Bostroem2003Assessing}
Bostr\"om, J., Greenwood, J.~R., and Gottfries, J. (2003).
\newblock Assessing the performance of omega with respect to retrieving
  bioactive conformations.
\newblock {\em J Mol Graph Model\/}, {\bf 21}(5), 449--462.

\bibitem[Catapano and Manji(2007)Catapano and Manji]{Catapano2007G}
Catapano, L.~A. and Manji, H.~K. (2007).
\newblock G protein-coupled receptors in major psychiatric disorders.
\newblock {\em Biochim Biophys Acta\/}, {\bf 1768}(4), 976--993.

\bibitem[Cavasotto {\em et~al.}(2003)Cavasotto, Orry, and
  Abagyan]{Cavasotto2003Structure-based}
Cavasotto, C.~N., Orry, A. J.~W., and Abagyan, R.~A. (2003).
\newblock Structure-based identification of binding sites, native ligands and
  potential inhibitors for g-protein coupled receptors.
\newblock {\em Proteins\/}, {\bf 51}(3), 423--433.

\bibitem[Chenna {\em et~al.}(2003)Chenna, Sugawara, Koike, Lopez, Gibson,
  Higgins, and Thompson]{Chenna2003Multiple}
Chenna, R., Sugawara, H., Koike, T., Lopez, R., Gibson, T.~J., Higgins, D.~G.,
  and Thompson, J.~D. (2003).
\newblock Multiple sequence alignment with the clustal series of programs.
\newblock {\em Nucleic Acids Res\/}, {\bf 31}(13), 3497--3500.

\bibitem[Cuturi and Vert(2005)Cuturi and Vert]{Cuturi2005context-tree}
Cuturi, M. and Vert, J.-P. (2005).
\newblock The context-tree kernel for strings.
\newblock {\em Neural {N}etwork.}, {\bf 18}(4), 1111--1123.

\bibitem[Deshpande and Penn(2006)Deshpande and Penn]{Deshpande2006Targeting}
Deshpande, D.~A. and Penn, R.~B. (2006).
\newblock Targeting g protein-coupled receptor signaling in asthma.
\newblock {\em Cell Signal\/}, {\bf 18}(12), 2105--2120.

\bibitem[Dobson and Doig(2005)Dobson and Doig]{Dobson2005Predicting}
Dobson, P. and Doig, A. (2005).
\newblock Predicting enzyme class from protein structure without alignments.
\newblock {\em J. {M}ol. {B}iol.}, {\bf 345}(1), 187--199.

\bibitem[Erhan {\em et~al.}(2006)Erhan, L'heureux, Yue, and
  Bengio]{Erhan2006Collaborative}
Erhan, D., L'heureux, P.-J., Yue, S.~Y., and Bengio, Y. (2006).
\newblock Collaborative filtering on a family of biological targets.
\newblock {\em J Chem Inf Model\/}, {\bf 46}(2), 626--635.

\bibitem[Evers and Klabunde(2005)Evers and Klabunde]{Evers2005Structure-based}
Evers, A. and Klabunde, T. (2005).
\newblock Structure-based drug discovery using gpcr homology modeling:
  successful virtual screening for antagonists of the alpha1a adrenergic
  receptor.
\newblock {\em J Med Chem\/}, {\bf 48}(4), 1088--1097.

\bibitem[Evgeniou {\em et~al.}(2005)Evgeniou, Micchelli, and
  Pontil]{Evgeniou2005Learning}
Evgeniou, T., Micchelli, C., and Pontil, M. (2005).
\newblock Learning multiple tasks with kernel methods.
\newblock {\em J. Mach. Learn. Res.}, {\bf 6}, 615--637.

\bibitem[Fredholm {\em et~al.}(2007)Fredholm, Hökfelt, and
  Milligan]{Fredholm2007G-protein-coupled}
Fredholm, B.~B., Hökfelt, T., and Milligan, G. (2007).
\newblock G-protein-coupled receptors: an update.
\newblock {\em Acta Physiol (Oxf)\/}, {\bf 190}(1), 3--7.

\bibitem[Frimurer {\em et~al.}(2005)Frimurer, Ulven, Elling, Gerlach, Kostenis,
  and H{\"o}gberg]{Frimurer2005physicogenetic}
Frimurer, T.~M., Ulven, T., Elling, C.~E., Gerlach, L.-O., Kostenis, E., and
  H{\"o}gberg, T. (2005).
\newblock A physicogenetic method to assign ligand-binding relationships
  between 7tm receptors.
\newblock {\em Bioorg. Med. Chem. Lett.}, {\bf 15}(16), 3707--3712.

\bibitem[G{\"a}rtner {\em et~al.}(2003)G{\"a}rtner, Flach, and
  Wrobel]{Gartner2003graph}
G{\"a}rtner, T., Flach, P., and Wrobel, S. (2003).
\newblock On graph kernels: hardness results and efficient alternatives.
\newblock In B.~Sch{\"o}lkopf and M.~Warmuth, editors, {\em Proceedings of the
  {S}ixteenth {A}nnual {C}onference on {C}omputational {L}earning {T}heory and
  the {S}eventh {A}nnual {W}orkshop on {K}ernel {M}achines\/}, volume 2777 of
  {\em Lecture Notes in Computer Science\/}, pages 129--143, Heidelberg.

\bibitem[Gasteiger and Engel(2003)Gasteiger and
  Engel]{Gasteiger2003Chemoinformatics}
Gasteiger, J. and Engel, T., editors (2003).
\newblock {\em Chemoinformatics : a {T}extbook\/}.
\newblock Wiley.

\bibitem[Henikoff and Henikoff(1992)Henikoff and Henikoff]{Henikoff1992Amino}
Henikoff, S. and Henikoff, J.~G. (1992).
\newblock Amino acid substitution matrices from protein blocks.
\newblock {\em Proc Natl Acad Sci U S A\/}, {\bf 89}(22), 10915--10919.

\bibitem[Hill(2006)Hill]{Hill2006G-protein-coupled}
Hill, S.~J. (2006).
\newblock G-protein-coupled receptors: past, present and future.
\newblock {\em Br J Pharmacol\/}, {\bf 147 Suppl 1}, S27--S37.

\bibitem[Hopkins and Groom(2002)Hopkins and Groom]{Hopkins2002druggable}
Hopkins, A.~L. and Groom, C.~R. (2002).
\newblock The druggable genome.
\newblock {\em Nat Rev Drug Discov\/}, {\bf 1}(9), 727--730.

\bibitem[Horn {\em et~al.}(2003)Horn, Bettler, Oliveira, Campagne, Cohen, and
  Vriend]{Horn2003GPCRDB}
Horn, F., Bettler, E., Oliveira, L., Campagne, F., Cohen, F.~E., and Vriend, G.
  (2003).
\newblock {GPCRDB information system for G protein-coupled receptors}.
\newblock {\em Nucl. Acids Res.}, {\bf 31}(1), 294--297.

\bibitem[Humphrey {\em et~al.}(1996)Humphrey, Dalke, and
  Schulten]{Humphrey1996VMD:}
Humphrey, W., Dalke, A., and Schulten, K. (1996).
\newblock Vmd: visual molecular dynamics.
\newblock {\em J Mol Graph\/}, {\bf 14}(1), 33--8, 27--8.

\bibitem[Jaakkola {\em et~al.}(2000)Jaakkola, Diekhans, and
  Haussler]{Jaakkola2000Discriminative}
Jaakkola, T., Diekhans, M., and Haussler, D. (2000).
\newblock A {D}iscriminative {F}ramework for {D}etecting {R}emote {P}rotein
  {H}omologies.
\newblock {\em J. {C}omput. {B}iol.}, {\bf 7}(1,2), 95--114.

\bibitem[Jacob and Vert(2008)Jacob and Vert]{Jacob2008Efficient}
Jacob, L. and Vert, J.-P. (2008).
\newblock {E}fficient peptide-{MHC}-{I} binding prediction for alleles with few
  known binders.
\newblock {\em Bioinformatics\/}.
\newblock To appear.

\bibitem[Jaroch and Weinmann(2006)Jaroch and Weinmann]{2006Chemical}
Jaroch, S.~E. and Weinmann, H., editors (2006).
\newblock {\em Chemical Genomics: Small Molecule Probes to Study Cellular
  Function\/}.
\newblock Ernst Schering Research Foundation Workshop. Springer.

\bibitem[Kashima {\em et~al.}(2003)Kashima, Tsuda, and
  Inokuchi]{Kashima2003Marginalized}
Kashima, H., Tsuda, K., and Inokuchi, A. (2003).
\newblock Marginalized {K}ernels between {L}abeled {G}raphs.
\newblock In T.~Faucett and N.~Mishra, editors, {\em Proceedings of the
  {T}wentieth {I}nternational {C}onference on {M}achine {L}earning\/}, pages
  321--328. AAAI Press.

\bibitem[Kashima {\em et~al.}(2004)Kashima, Tsuda, and
  Inokuchi]{Kashima2004Kernels}
Kashima, H., Tsuda, K., and Inokuchi, A. (2004).
\newblock Kernels for graphs.
\newblock In B.~Sch{\"o}lkopf, K.~Tsuda, and J.~Vert, editors, {\em Kernel
  {M}ethods in {C}omputational {B}iology\/}, pages 155--170. MIT Press.

\bibitem[Klabunde(2006)Klabunde]{Klabunde2006ChemogenomicsA}
Klabunde, T. (2006).
\newblock Chemogenomics approaches to ligand design.
\newblock In {\em Ligand Design for G Protein-coupled Receptors\/}, chapter~7,
  pages 115--135. Wiley-VCH.

\bibitem[Klabunde(2007)Klabunde]{Klabunde2007Chemogenomic}
Klabunde, T. (2007).
\newblock Chemogenomic approaches to drug discovery: similar receptors bind
  similar ligands.
\newblock {\em Br J Pharmacol\/}, {\bf 152}, 5--7.

\bibitem[Kratochwil {\em et~al.}(2005)Kratochwil, Malherbe, Lindemann, Ebeling,
  Hoener, Mühlemann, Porter, Stahl, and Gerber]{Kratochwil2005automated}
Kratochwil, N.~A., Malherbe, P., Lindemann, L., Ebeling, M., Hoener, M.~C.,
  Mühlemann, A., Porter, R. H.~P., Stahl, M., and Gerber, P.~R. (2005).
\newblock An automated system for the analysis of g protein-coupled receptor
  transmembrane binding pockets: alignment, receptor-based pharmacophores, and
  their application.
\newblock {\em J Chem Inf Model\/}, {\bf 45}(5), 1324--1336.

\bibitem[Kristiansen {\em et~al.}(1996)Kristiansen, Dahl, and
  Edvardsen]{Kristiansen1996database}
Kristiansen, K., Dahl, S.~G., and Edvardsen, O. (1996).
\newblock A database of mutants and effects of site-directed mutagenesis
  experiments on g protein-coupled receptors.
\newblock {\em Proteins\/}, {\bf 26}(1), 81--94.

\bibitem[Kuang {\em et~al.}(2005)Kuang, Ie, Wang, Wang, Siddiqi, Freund, and
  Leslie]{Kuang2005Profile-based}
Kuang, R., Ie, E., Wang, K., Wang, K., Siddiqi, M., Freund, Y., and Leslie, C.
  (2005).
\newblock Profile-based string kernels for remote homology detection and motif
  extraction.
\newblock {\em J Bioinform Comput Biol\/}, {\bf 3}(3), 527--550.

\bibitem[Kubinyi {\em et~al.}(2004)Kubinyi, M{\"u}ller, Mannhold, and
  Folkers]{Kubinyi2004Chemogenomics}
Kubinyi, H., M{\"u}ller, G., Mannhold, R., and Folkers, G., editors (2004).
\newblock {\em Chemogenomics in Drug Discovery: A Medicinal Chemistry
  Perspective\/}.
\newblock Methods and Principles in Medicinal Chemistry. Wiley-VCH.

\bibitem[Leslie {\em et~al.}(2002)Leslie, Eskin, and Noble]{Leslie2002spectrum}
Leslie, C., Eskin, E., and Noble, W. (2002).
\newblock The spectrum kernel: a string kernel for {SVM} protein
  classification.
\newblock In R.~B. Altman, A.~K. Dunker, L.~Hunter, K.~Lauerdale, and T.~E.
  Klein, editors, {\em Proceedings of the {P}acific {S}ymposium on
  {B}iocomputing 2002\/}, pages 564--575. World Scientific.

\bibitem[Leslie {\em et~al.}(2004)Leslie, Eskin, Cohen, Weston, and
  Noble]{Leslie2004Mismatch}
Leslie, C.~S., Eskin, E., Cohen, A., Weston, J., and Noble, W.~S. (2004).
\newblock Mismatch string kernels for discriminative protein classification.
\newblock {\em Bioinformatics\/}, {\bf 20}(4), 467--476.

\bibitem[Lin and Civelli(2004)Lin and Civelli]{Lin2004Orphan}
Lin, S. H.~S. and Civelli, O. (2004).
\newblock Orphan g protein-coupled receptors: targets for new therapeutic
  interventions.
\newblock {\em Ann Med\/}, {\bf 36}(3), 204--214.

\bibitem[Mah{\'e} {\em et~al.}(2005)Mah{\'e}, Ueda, Akutsu, Perret, and
  Vert]{Mahe2005Graph}
Mah{\'e}, P., Ueda, N., Akutsu, T., Perret, J.-L., and Vert, J.-P. (2005).
\newblock Graph kernels for molecular structure-activity relationship analysis
  with support vector machines.
\newblock {\em J {C}hem {I}nf {M}odel\/}, {\bf 45}(4), 939--51.

\bibitem[Mah{\'e} {\em et~al.}(2006)Mah{\'e}, Ralaivola, Stoven, and
  Vert]{Mahe2006Pharmacophore}
Mah{\'e}, P., Ralaivola, L., Stoven, V., and Vert, J.-P. (2006).
\newblock The pharmacophore kernel for virtual screening with support vector
  machines.
\newblock {\em J Chem Inf Model\/}, {\bf 46}(5), 2003--2014.

\bibitem[Okuno {\em et~al.}(2006)Okuno, Yang, Taneishi, Yabuuchi, and
  Tsujimoto]{Okuno2006GLIDA}
Okuno, Y., Yang, J., Taneishi, K., Yabuuchi, H., and Tsujimoto, G. (2006).
\newblock Glida: Gpcr-ligand database for chemical genomic drug discovery.
\newblock {\em Nucleic Acids Res\/}, {\bf 34}(Database issue), D673--D677.

\bibitem[Perola and Charifson(2004)Perola and
  Charifson]{Perola2004Conformational}
Perola, E. and Charifson, P.~S. (2004).
\newblock Conformational analysis of drug-like molecules bound to proteins: an
  extensive study of ligand reorganization upon binding.
\newblock {\em J Med Chem\/}, {\bf 47}(10), 2499--2510.

\bibitem[Qiu {\em et~al.}(2007)Qiu, Hue, Ben-Hur, Vert, and
  Noble]{Qiu2007structural}
Qiu, J., Hue, M., Ben-Hur, A., Vert, J.-P., and Noble, W.~S. (2007).
\newblock A structural alignment kernel for protein structures.
\newblock {\em Bioinformatics\/}, {\bf 23}(9), 1090--1098.

\bibitem[Ralaivola {\em et~al.}(2005)Ralaivola, Swamidass, Saigo, and
  Baldi]{Ralaivola2005Graph}
Ralaivola, L., Swamidass, S.~J., Saigo, H., and Baldi, P. (2005).
\newblock Graph kernels for chemical informatics.
\newblock {\em Neural {N}etw.}, {\bf 18}(8), 1093--1110.

\bibitem[Rognan(2007)Rognan]{Rognan2007Chemogenomic}
Rognan, D. (2007).
\newblock Chemogenomic approaches to rational drug design.
\newblock {\em Br J Pharmacol\/}, {\bf 152}, 38--52.

\bibitem[Rolland {\em et~al.}(2005)Rolland, Gozalbes, Nicolaï, Paugam, Coussy,
  Barbosa, Horvath, and Revah]{Rolland2005G-protein-coupled}
Rolland, C., Gozalbes, R., Nicolaï, E., Paugam, M.-F., Coussy, L., Barbosa, F.,
  Horvath, D., and Revah, F. (2005).
\newblock G-protein-coupled receptor affinity prediction based on the use of a
  profiling dataset: Qsar design, synthesis, and experimental validation.
\newblock {\em J Med Chem\/}, {\bf 48}(21), 6563--6574.

\bibitem[Russell and Barton(1992)Russell and Barton]{Russell1992Multiple}
Russell, R.~B. and Barton, G.~J. (1992).
\newblock Multiple protein sequence alignment from tertiary structure
  comparison: assignment of global and residue confidence levels.
\newblock {\em Proteins\/}, {\bf 14}(2), 309--323.

\bibitem[Sch{\"o}lkopf {\em et~al.}(2004)Sch{\"o}lkopf, Tsuda, and
  Vert]{Schoelkopf2004Kernel}
Sch{\"o}lkopf, B., Tsuda, K., and Vert, J.-P. (2004).
\newblock {\em Kernel {M}ethods in {C}omputational {B}iology\/}.
\newblock MIT Press.

\bibitem[Todeschini and Consonni(2002)Todeschini and
  Consonni]{Todeschini2002Handbook}
Todeschini, R. and Consonni, V. (2002).
\newblock {\em Handbook of Molecular Descriptors\/}.
\newblock Wiley-VCH.

\bibitem[Tsuda {\em et~al.}(2002)Tsuda, Kin, and Asai]{Tsuda2002Marginalized}
Tsuda, K., Kin, T., and Asai, K. (2002).
\newblock Marginalized {K}ernels for {B}iological {S}equences.
\newblock {\em Bioinformatics\/}, {\bf 18}, S268--S275.

\bibitem[Vert(2002)Vert]{Vert2002tree}
Vert, J.-P. (2002).
\newblock A tree kernel to analyze phylogenetic profiles.
\newblock {\em Bioinformatics\/}, {\bf 18}, S276--S284.

\bibitem[Vert {\em et~al.}(2004)Vert, Saigo, and Akutsu]{Vert2004Local}
Vert, J.-P., Saigo, H., and Akutsu, T. (2004).
\newblock Local alignment kernels for biological sequences.
\newblock In B.~Sch{\"o}lkopf, K.~Tsuda, and J.~Vert, editors, {\em Kernel
  {M}ethods in {C}omputational {B}iology\/}, pages 131--154. MIT Press.

\bibitem[Vert {\em et~al.}(2006)Vert, Bach, and Evgeniou]{Vert2006Low-rank}
Vert, J.-P., Bach, F., and Evgeniou, T. (2006).
\newblock Low-rank matrix factorization with attributes.

\end{thebibliography}
\end{document}